\documentclass{article}

\usepackage{amsmath}
\usepackage{wrapfig}
\usepackage{graphicx}
\usepackage{color}
\usepackage[numbers]{natbib}

 \usepackage[preprint]{nips_2018}

\usepackage[utf8]{inputenc} % allow utf-8 input
\usepackage[T1]{fontenc}    % use 8-bit T1 fonts
\usepackage{hyperref}       % hyperlinks
\usepackage{url}            % simple URL typesetting
\usepackage{booktabs}       % professional-quality tables
\usepackage{amsfonts}       % blackboard math symbols
\usepackage{nicefrac}       % compact symbols for 1/2, etc.
\usepackage{microtype}      % microtypography

\title{Winner-Take-All as Basic Probabilistic Inference Unit of Neuronal Circuits}

\author{
  Zhaofei Yu\\
  School of EE\&CS\\
  Peking University\\
  Beijing, China \\
  \texttt{yuzf12@pku.edu.cn} \\
  \And
    Yonghong Tian\\
  School of EE\&CS\\
  Peking University\\
  Beijing, China \\
  \texttt{yhtian@pku.edu.cn} \\
  \And
    Tiejun Huang\\
  School of EE\&CS\\
 Peking University\\
 Beijing, China \\
 \texttt{tjhuang@pku.edu.cn} \\ 
   \And
 Jian K. Liu\\
 Center of Systems Neuroscience\\ University of Leicester\\
 Leicester LE1 7HA, UK \\
 \texttt{jian.liu@leicester.ac.uk} \\ 
}

\begin{document}
% \nipsfinalcopy is no longer used

\maketitle

\begin{abstract}
Experimental observations of neuroscience suggest that the brain is working a probabilistic way when computing information with uncertainty. This processing could be modeled as Bayesian inference. However, it remains unclear how Bayesian inference could be implemented at the level of neuronal circuits of the brain. In this study, we propose a novel general-purpose neural implementation of probabilistic inference based on a ubiquitous network of cortical microcircuits, termed winner-take-all (WTA) circuit. We show that each WTA circuit could encode the distribution of states defined on a variable. By connecting multiple WTA circuits together, the joint distribution can be represented for arbitrary probabilistic graphical models. Moreover, we prove that the neural dynamics of WTA circuit is able to implement one of the most powerful inference methods in probabilistic graphical models, mean-field inference. We show that the synaptic drive of each spiking neuron in the WTA circuit encodes the marginal probability of the variable in each state, and the firing probability (or firing rate) of each neuron is proportional to the marginal probability.   Theoretical analysis and experimental results demonstrate that the WTA circuits can get comparable inference result as mean-field approximation. Taken together, our results suggest that the WTA circuit could be seen as the minimal inference unit of neuronal circuits.

\end{abstract}

\section{Introduction}
Humans are able to process information in the face of uncertainty in the sensory, motor and cognitive domains \cite{meyniel2015confidence}. Confidence in any decision-making tasks does not come from the one-time judgment of all of the uncertainties.  Instead, we estimate the uncertainty and inference the problem by cumulating our knowledge based on some rules of probabilistic inference. The processes like these can be understood as Bayesian inference. There is an increasing volume of behavioral and physiological evidence that human and monkeys (and other animals) can represent probabilities and implement probabilistic computations in a fashion of Bayesian inference with some types of neuronal and circuitry mechanisms in the brain \cite{ernst2002humans,kording2004bayesian, Pouget2013Probabilistic}. Despite this, it remains unclear how probabilistic inference is implemented by our neuronal system. For a perspective of theoretical consideration, the question is how to implement such a probabilistic inference with a network of spiking neurons. 

A few models have been proposed to relate the dynamics of spiking neural networks to the inference equations of Belief Propagation (BP) algorithm, an inference algorithm commonly used in probabilistic graphical model, which is exact on acyclic graphical models and approximate on cyclic graphical models \cite{rao2004bayesian,Sch2006The,Steimer2009Belief,Litvak2009Cortical,George2009Towards,Friston2017Active}. All these studies try to prove that the dynamics of spiking neural networks could implement basic computation of BP algorithm, thus the time-course of the dynamics of spiking neural networks can be understood as the inference process. A recent approach is implementing tree-based reparameterization algorithm \cite{Raju2016Inference}, of which BP is a special case that only considers the reparameterization over just two eighboring nodes and their corresponding edge. 
In summary, most of the previous studies require each neuron and synapse conduct complicated computation, thus they are hard to be generalized to a general framework. In fact in the brain, a
single neuron or a group of neurons should work in a relatively simple style while complex functions could be achieved when they are combined together. This could be achieved if there is a basic computation motif in the neuronal circuit, and then a combination of them can move towards to the complex functions. 
Therefore, it is worth considering what could be the basic inference motif in our neuronal system. If so, can the composition of these basic motifs implement inference for any Bayesian model with multiple layers and arbitrary scale?

Out of these possible motifs, there is one type of motif that has been intensively studied from the theoretical viewpoint in the last decades. It is named as winner-take-all (WTA) circuit that a microcircuit with an ensemble of excitatory cells with lateral inhibition as suggested by experimental observations \cite{okun2008instantaneous, douglas2004neuronal}. Within this network motif, the competition between excitatory cells induced by the inhibition makes the WTA circuit suitable for implementation of many types of neuronal computations, such as feature selection, attention and decision making \cite{douglas2004neuronal, carandini2012normalization, itti2001computational}.

However, the functional importance of a large scale of neuronal circuit with abundant WTA circuits remains unknown. Especially, from a computational perspective, it is unclear if probabilistic inference can be emerged with a combination of WTA circuits. 

In this paper, we show that each WTA circuit can encode the state of a variable in a probabilistic graph model (PGM), the combination of which could represent the joint distribution defined on any PGM with synaptic weights and input current encoding the potential functions. Moreover, we prove that the neural dynamics of a network consisted of multiple WTA circuits is exactly equivalent to mean-field inference algorithm of probabilistic graphical models. We show the synaptic drive of each spiking neuron in the WTA circuit encodes marginal probability of the variable being in each state, and the firing probability (or firing rate) of each neuron is proportional to the marginal probability of the variable being in each state. Our results suggest that the WTA circuit can be seen as the minimal inference motif of neuronal system.

\section{Inference by Mean-Field Approximation}

In this section, we first briefly review probabilistic graphical model and variational inference algorithm with mean-field approximation, and then drive the differential equation which is equivalent to the mean-field inference algorithm.   

\subsection{Probabilistic Graphical Model}

\begin{figure}
  \begin{minipage}[c]{0.57\textwidth}
    \includegraphics[width=9cm]{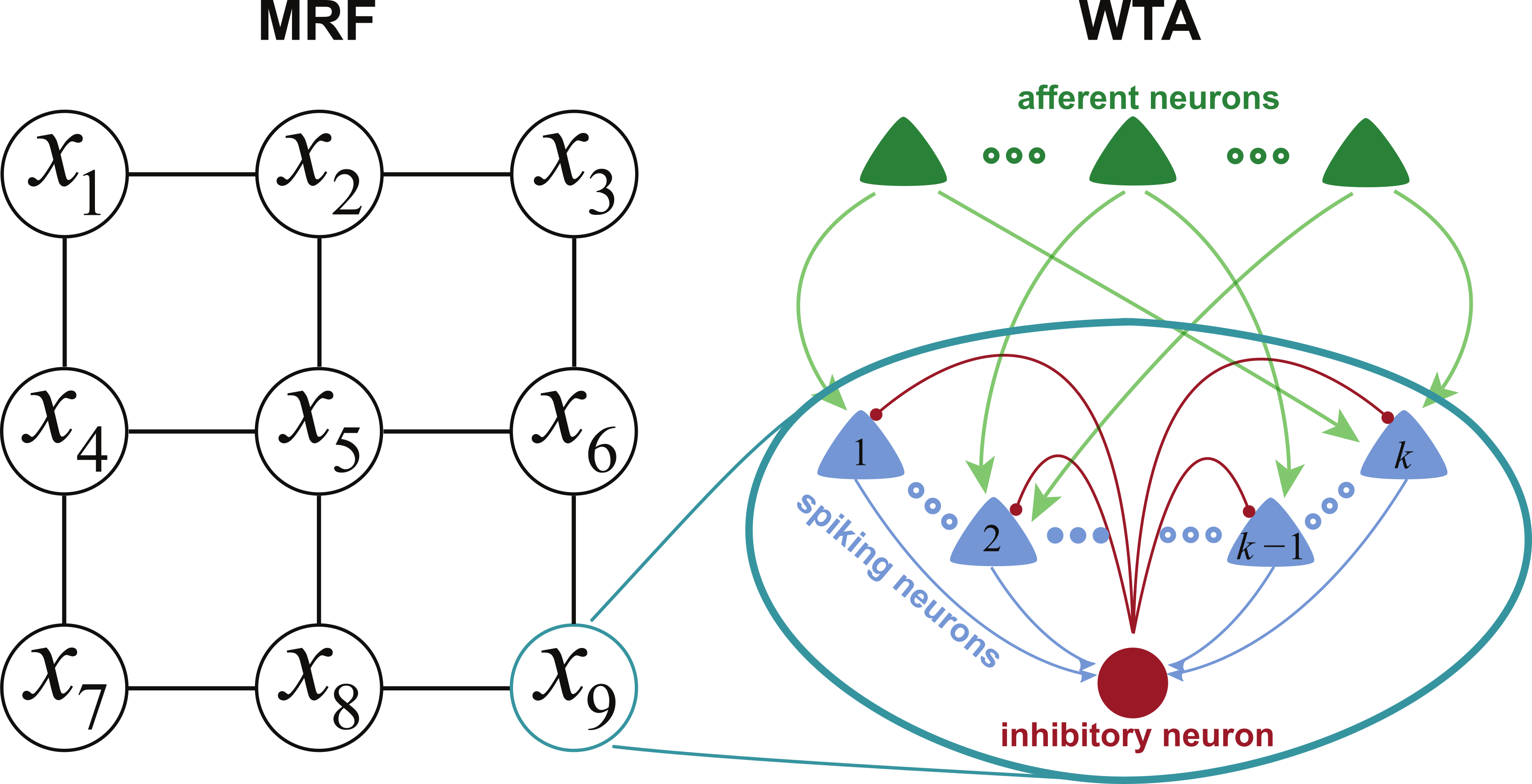}
  \end{minipage}\hfill
  \begin{minipage}[c]{0.35\textwidth}
    \caption{Illustration of MRF and WTA circuit. A variable of $k$ states in MRF can be presented by $k$ output excitatory neurons of WTA circuit. Competition mechanism of WTA is achieved by excitatory neurons (blue) and one inhibitory neuron (red). Each excitatory neuron receives the input current represented by afferent neurons, which encode the potential functions defined on each node of MRF.
    } \label{fig:model}
  \end{minipage}
\end{figure}

Probabilistic graphical model (PGM) provides a powerful formalism for multivariate statistical modeling by combining graph theory and probability theory \cite{koller2009probabilistic}. It has been widely used in computer vision, signal processing, and computational neuroscience. In computational neuroscience, PGMs are used to model the inference process of the human brain. In this study, we only focus on the inference of undirected probabilistic graphical model, also known as Markov random fields (MRFs). The results can be easily generalized to directed probabilistic graphical models due to the fact that a directed probabilistic graphical model can be converted to an undirected probabilistic graphical model with moralization \cite{koller2009probabilistic, Jordan1999An}. As shown in Fig. \ref{fig:model}, the joint distribution $p(x_1,x_2,...,x_n)$ over $n$ variables, $\textbf{x}=\left \{ x_1,x_2,...,x_n\right\}$ defined on a MRF can be factorized into a product of potential functions according to the graph structure, that is,
\begin{equation}
\label{pgm1}
p(x_1,x_2,...,x_n)=\frac{1}{Z}\prod_{(i,j)\in E}  \Psi_{ij}(x_i,x_j) \prod_{(i)\in V}\Psi_i (x_i),
\end{equation}  
where $E$ is the set of edges and $V$ is the set of nodes, $\Psi_{ij}(x_i,x_j)$ and $\Psi_i (x_i)$ represent the potential functions of each edge and node respectively. $Z$ is a normalized constant, which equals $\sum_{x_1,x_2,...,x_n} \prod_{(i,j)\in E}  \Psi_{ij}(x_i,x_j) \prod_{(i)\in V}\Psi_i (x_i)$. If one defines that $\theta_{ij}(x_i,x_j)=\ln\Psi_{ij}(x_i,x_j)$ and $\theta_{i}(x_i)=\ln\Psi_{i}(x_i)$, equation \eqref{pgm1} can be reformulated as:
\begin{equation}
p(x_1,x_2,...,x_n)=\frac{1}{Z} \exp \left(\sum_{(i,j)\in E}\theta_{ij}(x_i,x_j)+ \sum_{(i)\in V}\theta_{i}(x_i)\right).
\end{equation}

\subsection{Variational Inference with Mean-Field Approximation}

When modeling the inference process of the human brain with graphical models, the inference problem includes two folds: (1) computing marginal distribution of each variable $x_i$, that is $p(x_i)=\sum_{\textbf{x} \backslash  x_i} p(x_1,x_2,...,x_n)$, where ${\textbf{x} \backslash  x_i}$ represents all the variables in $\textbf{x}$ expect variable $x_i$, (2) computing posterior probability $p(x_i|x_j)$. In fact these two problems are naturally coupled as $p(x_i|x_j)=\frac{\sum_{\textbf{x} \backslash  x_i,x_j} p(x_1,x_2,...,x_n)}{\sum_{\textbf{x} \backslash  x_j} p(x_1,x_2,...,x_n)}$, thus we only need to consider marginal inference. 
As exact inference of MRF is an NP-complete problem, people often use
efficient variational approximate inference algorithms,
the idea of which is converting an inference problem to the optimization problem $\min_{q(\textbf{x})} \text{KL} \left(q(\textbf{x})||p(\textbf{x})\right)$.
Here the initial distribution $p(\textbf{x})$ is approximated by a distribution $q(\textbf{x})$ which belongs to a family of tractable distributions, $\text{KL}(\cdot)$ represents the Kullback-Leibler divergence.
The mean-filed approximation is obtained by setting $q(\textbf{x})$ to be a fully factorized distribution, that is, 
$q(\textbf{x})=\prod_i {m_i(x_i)}$. By constraining $\sum_{x_i} m_i(x_i)=1$ and differentiating $\text{KL} \left(q(\textbf{x})||p(\textbf{x})\right)$ with respect to $m_i(x_i)$, one can get the following mean-field inference equations
\begin{equation}
\label{meanfileddis}
m_i^{\lambda+1}(x_i)=\frac{\exp \left(\sum_{j\in N(i)} \sum_{x_j} \theta_{ij}(x_i,x_j) m_j^\lambda (x_j)+ \theta_{i}(x_i) \right)}{\sum_{x_i} \exp \left(\sum_{j\in N(i)} \sum_{x_j} \theta_{ij}(x_i,x_j) m_j^\lambda (x_j)+ \theta_{i}(x_i) \right)}, i=1,2,...n,
\end{equation}
where $\lambda$ denotes the number of iterations, $m_i^{\lambda}(x_i)$ represents the information received by node $i$ (approximate marginal probability of variable $x_i$) in the $\lambda$ th iteration. When all the messages converge to the fixed points, the marginal probability $p(x_i)$ can be approximated by the steady-state $m_i^{\infty}(x_i)$.
It is easy to prove that the following differential equation has the same fixed point as equation \eqref{meanfileddis}:
\begin{equation}
\label{meanfiledcon}
\tau_0 \frac{d m_{i}^k(t)}{dt}=-m_{i}^k(t)+\frac{\exp \left(\sum_{j\in N(i)} \sum_{l =1}^{X_j} \theta_{ij}^{kl} m_{j}^l(t)+ \theta_{i}^k \right)}{\sum_{k} \exp \left(\sum_{j\in N(i)} \sum_{l =1}^{X_j} \theta_{ij}^{kl} m_{j}^l(t)+ \theta_{i}^k \right)},
\begin{array}{*{20}{c}}
{~i=1,2,...n}\\
{~k=1,2,...X_i}
\end{array},
\end{equation}
where $\tau_0$ is a time constant, $X_i$ and $X_j$ denote the number of all possible states of variables $x_i$ and $x_j$ respectively. Note that we have written $m_i^{t}(x_i=k)$ as $m_{i}^k(t)$, $\theta_{ij}(x_i=k,x_j=l)$ as $\theta_{ij}^{kl}$, and $\theta_{i}(x_i=k)$ as $\theta_{i}^k$ for notational convenience.

\section{Spiking Neural Network}

\subsection{Spiking Neuron Model}

In this study, the spiking neuron is modeled by a standard stochastic variant of the spike response model \cite{gerstner2014neuronal}, which is a generalization of the
leaky integrate-and-fire neuron. Considering a network of $K$ spiking neurons $z_1,z_2,...,z_K$, the output spike train of neuron $z_k$ is denoted by $S_k(t)$, which is defined as a sum of Dirac delta pulses positioned at the spike times $t_k^{(1)}, t_k^{(2)}, \dots$, i.e., $S_k(t) = \sum_l \delta(t-t_k^{(l)})$. It's obvious to see  $S_k(t)=1$ if neuron $z_k$ fires at time $t$ and $S_k(t)=0$ otherwise. In this model, the membrane potential $u_k(t)$ of neuron $z_k$ at time $t$ is given by:
\begin{equation}
\label{onemem}
u_k(t)=\sum_{j\in \text{pre}(k)}  w_{kj} \int_{0}^{t} \kappa (t-s)  S_j(s)ds~+~I_k(t),
\end{equation} 
where $\text{pre}(k)$ denotes the set of pre-synaptic neurons for neuron $z_k$, %$I_j(t)$ represents the input current to neuron $z_k$ from pre-synaptic neuron $z_j$,
$I_k(t)$ represents the input current from outside stimulus.  
$w_{kj}$ denotes the synaptic weight between neuron $z_j$ and $z_k$, $\kappa (t)$ describes the voltage response to a short current pulse. Here we use the standard exponential kernel $\kappa (t)$ as in  \cite{fremaux2010functional}: 
\begin{equation}
\label{filter}
\kappa (t) = \frac{1}{\tau} \exp \left(-\frac{t}{\tau}\right),
\end{equation}
with the membrane time constant $\tau$. In the standard stochastic variant of the
spike response model, the strict firing threshold of membrane potential is replaced by a noisy threshold, which means that a neuron can fire  stochastically at any membrane potential \cite{gerstner2014neuronal}.
To be specific, neuron $z_k$ fires a spike at time $t$ with an instantaneous probability $\rho_k(t)$, which is often modeled by an exponential function of the membrane potential:
\begin{equation}
\rho_k(t)=\rho \exp \left(u_k(t)-\theta\right),
\end{equation}
where $\theta$ decides the firing threshold and $\rho$ scales the firing rate of the neuron $z_k$. One can find that the firing rate increases as the distance between membrane potential and firing threshold decreases. It also has been shown that this model is in good agreement with real neurons \cite{jolivet2006predicting}. 

\subsection{Winner-Take-All Circuit}

Winner-take-all (WTA) circuit has been suggested as an ubiquitous motif of cortical microcircuits \cite{douglas2004neuronal}. We consider a WTA circuit of $K$ output spiking neurons and an
inhibitory neuron as in Fig. \ref{fig:model}. The output spiking neurons $z_1, z_2,..., z_K$ mutually inhibit each other through the inhibitory neuron. Thus, all the neurons in the output layer
are in competition against each other so that they cannot fire
simultaneously.

In this study, we consider the WTA model used in \cite{nessler2013bayesian,kappel2014stdp}, where all neurons are allowed to fire with non-zero probability. Considering all the neurons in a WTA circuit are subject to the same lateral inhibition, the firing probability of neuron
$z_k$ in the WTA circuit at time $t$ is determined by \cite{nessler2013bayesian}:
\begin{equation}
\label{WTA_fire}
\rho_k(t)=\frac{\rho }{Q(t)} \exp \left(u_k(t)-\theta\right),
\end{equation}
where $\rho$ scales the firing rate of neurons. $Q(t)$ represents the
divisive inhibition between the neurons in the WTA circuit, and is defined as $Q(t)=\sum_k \exp(u_k(t)-\theta)$. Thus, equation \eqref{WTA_fire} can be rewritten as:
\begin{equation}
\label{WTA_firenew}
\rho_k(t)=\frac{\rho }{\sum_k \exp(u_k(t)-\theta)} \exp \left(u_k(t)-\theta\right)= \rho \frac{\exp \left(u_k(t)\right)}{\sum_k \exp(u_k(t))}.
\end{equation}
This WTA circuit works like a soft-max function. At each
time, all neurons can fire with non-zero probability, but the
neuron with the highest membrane potential has the highest firing
probability.

\section{Neural Implementation of Mean-Field Inference}

\begin{figure}
	\begin{center}
		\includegraphics[width=\textwidth]{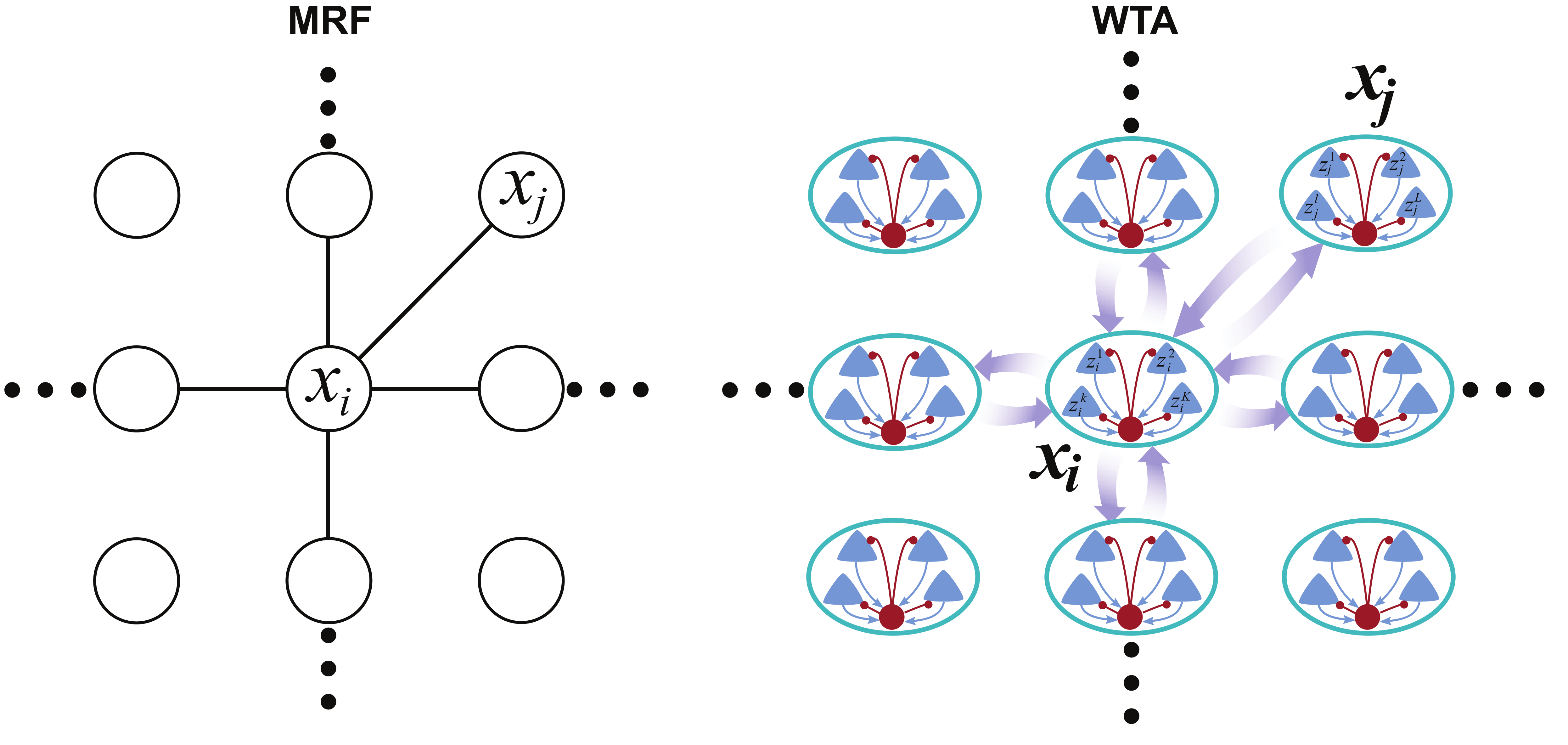}
	\end{center}
	\caption{MRF network represented by a combination of WTA circuits. (Left) MRF network with multiple variables, where $x_i$ and $x_j$ are connected. (Right) Corresponding neural network. The network consists of multiple WTA circuits, where each neuron encodes one state of a variable, and each WTA circuit encodes the distribution defined on the variable. When two nodes ($x_i$ and $x_j$) are connected in the MRF, all the spiking output neurons in the corresponding WTA circuit are fully connected  (connections between the $i$th WTA circuit and the $j$th circuit). Here the synaptic weights and input currents encode the potential functions defined on the edges and nodes respectively. The whole circuit is able to encode the joint distribution defined on the MRF.  
	} \label{fig:structure}
\end{figure}

In this section, we first show how a basic WTA circuit encode the distribution of a variable defined on a node and how a network consisted of WTA circuits represent the joint distribution defined on a MRF. Then we prove that there exists an exact equivalence between the mean-field inference equation of  a MRF and the dynamic equation of a spiking neural  network consisted of WTA circuits.

\subsection{Representation of Distributions with WTA Circuits}

In order to enable the combination of WTA circuits to implement arbitrary inference of MRFs, one needs to specify how the assignment $(x_1,x_2,...,x_n) \in \{1,2,...,X_1\}\times\{1,2,...,X_2\}...\times\{1,2,...,X_n\}$ of values to these variables defined on MRFs can be represented by the spiking activities of WTA circuits, where $X_1,X_2,...,X_n$ represents the number of states of variables $x_1,x_2,...,x_n$ respectively. In fact, each WTA circuit can represent the states of a variable in a MRF. To be specific, consider a WTA circuit with $X_i$ output spiking neurons $z_1,z_2,...,z_{X_i}$, we say that a variable $x_i$ is represented by the firing activity of a WTA circuit at time $t$ if:
\begin{equation}
x_i=k \Leftrightarrow \text{neuron}~z_k~ \text{fires at time}~t~~(k=1,2,...,X_i).
\end{equation}
In this way, each neuron $z_k (k=1,2,...,X_i)$ represents one of the $K$ possible values of variable $x_i$. 
If the firing probability of each output spiking neuron equals the probability of the variable being in each state, that is, $p(x_i=k)=p(z_k(t)=1),k=1,2,...,X_i$, then the firing activities of the WTA circuit encodes the distribution defined on the variable. 
One can read out the distribution 
by counting spikes from each neuron within a behaviorally
relevant time window of a few hundred milliseconds, which
is similar to the experimental results of monkey cortex\cite{yang2007probabilistic,gold2007neural}. Similarly, a spiking neural networks consisted of  $n$ WTA circuit can encode the joint distribution of $p(x_1,x_2,...,x_n)$ over $n$ variables $\textbf{x}=\left \{ x_1,x_2,...,x_n\right\}$ defined on a MRF.

\subsection{Network Architecture}

Here we illustrate the neural network architecture of WTA circuits to perform the inference of arbitrary MRF. Considering a MRF and its corresponding spiking neural network in Fig. \ref{fig:structure}. The neural network is composed of several WTA circuits, of which each WTA circuit represents (or encodes) a variable defined on each node. If there exists a connection between two nodes in the MRF, the output spiking neurons in the corresponding  WTA circuits are fully connected. The connection weights are used to encode the potential functions defined on the edges between the adjacent nodes in MRF. Also, each neuron can receive the input current from output stimulus (not shown in Fig. \ref{fig:structure}), which encodes the potential functions defined on each node of MRF.

\subsection{Spike-based Mean-field Inference in WTA circuits }

In order to prove the combination of WTA circuits can implement inference for arbitrary MRFs, here we prove that there exists an equivalence between the dynamic equation of WTA circuits and the differential equation \eqref{meanfiledcon} of the MRF as in Fig. \ref{fig:structure}.

Considering the spiking neural network in Fig. \ref{fig:structure}, there are $n$ WTA circuits. The $i$ th WTA circuit consists of $X_i$ output spiking neurons, which encodes the variable $x_i$. The $k$ th neuron in $i$ th WTA circuit is denoted as $z_i^k(t)$, which receives stimulus current $I_i^k(t)$ (not shown in Fig. \ref{fig:structure}) and synaptic inputs from the neurons in the neighboring WTA circuits. 
We denote the output spike train of neuron $z_i^k$ by $S_i^k(t)$, which is defined as a sum of Dirac delta pulses positioned at the spike times $t_i^{k,(1)}, t_i^{k,(2)}, \dots$, i.e., $S_i^k(t) = \sum_l \delta(t-t_i^{k,(l)})$. We assume that $\alpha_i^k(t)=\frac{1}{\rho}\int_{0}^{t} \frac{1}{\tau} \exp (-\frac{t-s}{\tau})  \rho_i^k(s)ds$, where $\alpha_i^k(t)$ is called synaptic drive \cite{dayan2001theoretical}, $\rho_i^k(s)$ represents the firing probability of the $k$ th neuron in $i$ th WTA circuit at time $s$, and $\rho$ has been defined in \eqref{WTA_fire} to scale the firing rate of neurons. If we take the derivative of $\alpha_i^k(t)$ with respect to time $t$, we obtain: 
\begin{equation}
\label{synapticdrive1}
\frac{d\alpha_i^k(t)}{dt}=\frac{1}{\tau \rho} \left(-\rho\alpha_i^k(t)+\rho_i^k(t)\right).
\end{equation}
 As neuron $z_i^k$ is in the $i$ th WTA circuit, the firing probability $\rho_i^k(t)$ of neuron $z_i^k$ equals $\rho \frac{\exp \left(u_i^k(t)\right)}{\sum_k \exp \left(u_i^k(t)\right)}$. According to equation \eqref{WTA_firenew}, thus equation \eqref{synapticdrive1} can be rewritten as:
\begin{align}
\label{synapticdrive1new}
\tau \frac{d\alpha_i^k(t)}{dt}=-\alpha_i^k(t)+\frac{\exp \left(u_i^k(t)\right)}{\sum_k \exp \left(u_i^k(t)\right)}.
\end{align}
According to equation \eqref{onemem}, the membrane potential $u_i^k$ of neuron  $z_i^k$ at time $t$ equals 
\begin{align}
\label{membrane1}
u_i^k(t)&=\sum_{j\in N(i)} \sum_{l=1}^{X_j} w_{ij}^{kl} \int_{0}^{t} \kappa (t-s)  S_j^l(s)ds+  I_i^k(t) \nonumber \\
&\approx \sum_{j\in N(i)} \sum_{l=1}^{X_j} w_{ij}^{kl} \int_{0}^{t} \frac{1}{\tau} \exp (-\frac{t-s}{\tau})   \rho_j^l(s)ds+  I_i^k(t)\nonumber \\
&=\sum_{j\in N(i)} \sum_{l=1}^{X_j} \rho w_{ij}^{kl}\alpha_j^l(t)+  I_i^k(t).
\end{align} 
where $N(i)$ represents all neighboring WTA circuits of the $i$ th WTA circuit, $w_{ij}^{km}$ denotes the synaptic weight between neuron $z_j^m$ and $z_i^k$. %$\xi_i^k$ denotes feedforward synaptic weight to neuron $z_i^k$. The equality holds as $\int_{0}^{t} \kappa (t-s) ds=1$. 
Note that here the output spike train $S_j^l(s)$ of neuron $z_j^l$ at time $s$ is approximated by the firing probability function $\rho_j^l(s)$, which is also used in \cite{dayan2001theoretical} when driving the dynamic equation of recurrent neural networks. By substituting equation \eqref{membrane1} into equation \eqref{synapticdrive1new}, one can get:
\begin{align}
\label{synaptivcon}
\tau \frac{d\alpha_i^k(t)}{dt}=-\alpha_i^k(t)+\frac{\exp \left(\sum_{j\in N(i)} \sum_{l=1}^{X_j} \rho w_{ij}^{kl}\alpha_j^l(t) +I_i^k\right)}{\sum_k \exp \left(\sum_{j\in N(i)} \sum_{l=1}^{X_j} \rho w_{ij}^{kl}\alpha_j^l(t) +I_i^k\right)}.
\end{align} 
Now one can find that a spiking neural network consisted of WTA circuits governed by \eqref{synaptivcon} can implement equation \eqref{meanfiledcon} if the following equations hold.
\begin{equation}
\label{equals1}
\tau=\tau_0, ~~~~~\rho w_{ij}^{kl}=\theta_{ij}^{kl},~~~~~ I_i^k=\theta_{i}^k,~~~~~ \alpha_i^k(t)=m_{i}^k(t).
\end{equation}
It means if the synaptic weights $w_{ij}^{kl}$ and input current $I_i^k$ encodes the potential functions $\theta_{ij}^{k}$ and $\theta_{i}^k$ respectively, then the synaptic drive of each spiking neuron in the WTA circuits equals marginal probability of the variable being in each state.
Note that when equation \eqref{synapticdrive1} and \eqref{synaptivcon} converges to the fix point, we have $\alpha_i^k(t)=\frac{\rho_i^k(t)}{\rho}$. Thus the firing probability (or firing rate) of each neuron is proportional to the marginal probability of the variable being in each state. Moreover,
the time course of neural firing rate can implement marginal
inference of MRFs. One can read out the inference result
by counting spikes from each neuron within a behaviorally
relevant time window of a few hundred milliseconds. If fact, the computation of WTA circuits in simulations can converge to the inference result very fast (see Supplementary Fig.~1). 

The proposed theory is working in continuous time domain. It is easy to converge it into a version where the dynamics of WTA is discrete in time for numerical simulation. Based on the equivalence derived above, the changes of firing probability in the discrete time bins can be seen as one iteration of mean-field inference equation \eqref{meanfileddis} (see Supplementary Materials).

\section{Simulations}

\begin{figure}
	\begin{center}
		\includegraphics[width=\textwidth]{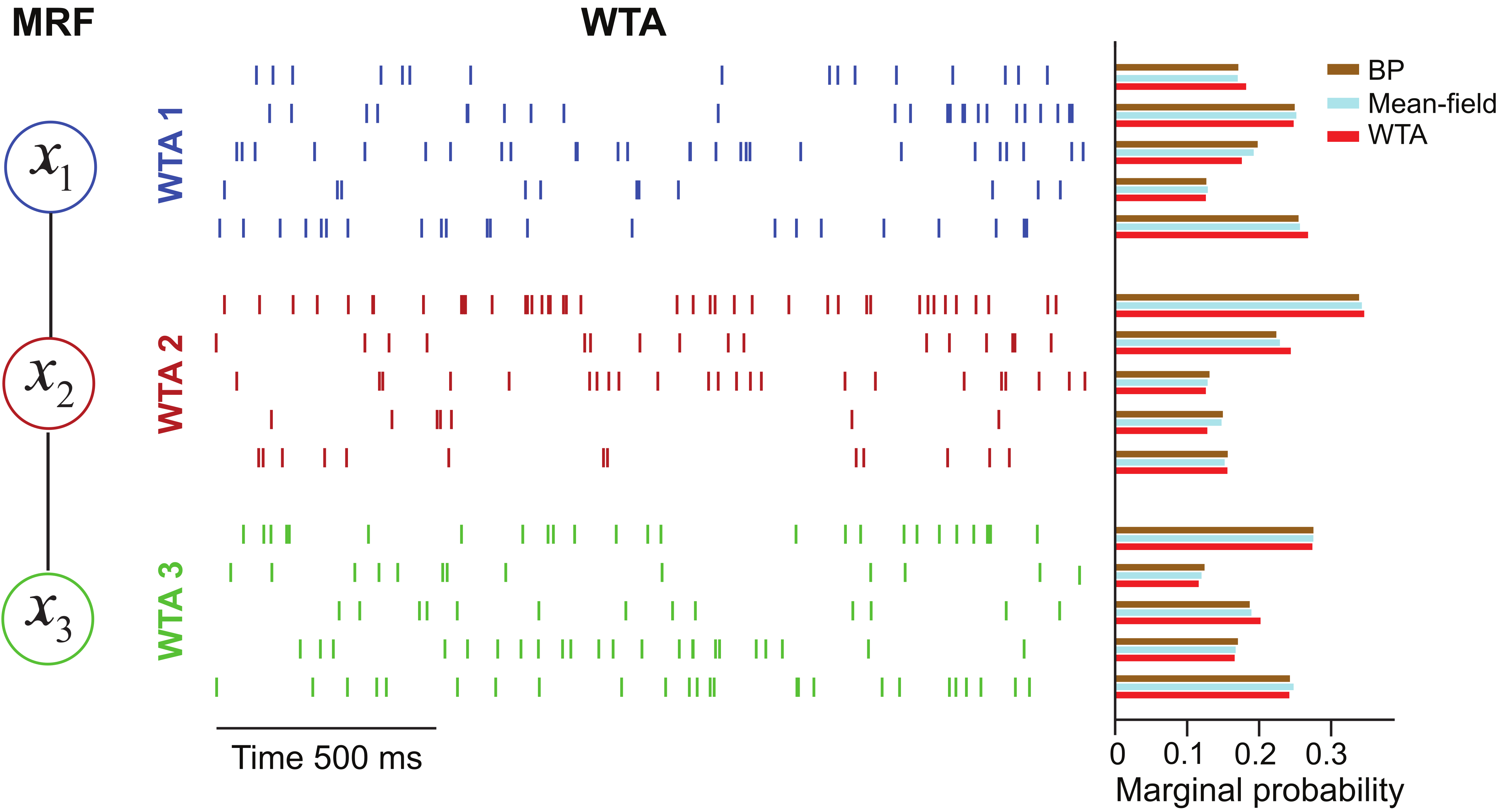}
	\end{center}
	\caption{Mean-field inference simulated with WTA circuits. (Left) A chain MRF network with three nodes. (Middle) Spiking activity of 15 neurons in three WTA circuits, where each WTA circuit has 5 neurons. All neurons are firing through the competition of WTA mechanism. (Right) Tight match of three difference inference methods for marginal probability: belief propagation (brown), mean-field inference (light blue), WTA spiking neural network (red). 
	} \label{fig:result1}
\end{figure}

To validate our computational framework, we evaluate the performance of WTA circuits through two simulations. Firstly, we present the comparison of WTA circuits with mean-field algorithm and belief propagation algorithm on a chain MRF with 3 nodes (shown in Fig. \ref{fig:result1}). Note that belief propagation can conduct exact inference for this MRF. We suppose that each node has 5 states, the potential functions
$\theta_{ij}(x_i,x_j)$ and $\theta_{i}(x_j)$ defined on each edge and node are created by randomly generating numbers from a
uniform distribution on $[0,1]$. A spiking neural network composed of 3 WTA circuits is used to implement inference, of which each WTA circuit includes 5 neurons. The synaptic weights between neurons and input currents are set according to equation \eqref{equals1}. In our simulation, the firing rate scaling factor $\rho$ is assumed to be 50Hz, thus we are able to map the firing rate $[0, 50]$Hz of each neuron to the probability $[0, 1]$ of each state of a variable. The firing activity of 15 neurons is shown in Fig. \ref{fig:result1}, where one can find all the neurons can fire. 

The performance of inference is shown as the histograms of the firing rate of 15 neurons 
%in a 2-s window 
in Fig. \ref{fig:result1}, where we compare the inference result of the MRF with different methods. As a result, the tight match between the three inference algorithms suggests that the WTA spiking neural network can perform mean-field variational inference with high accuracy. We also use the relative error $\frac{1}{n} \sum_{i=1}^n \| p(x_i)- \rho_i \|  / \|p(x_i)\|$ to evaluate the divergence between the marginal probability $p(x_i)$ ($i=1,2,...n$) obtained by belief propagation and mean firing probability $\rho_i$ ($i=1,2,...n$) of each neuron. The relative error decreases over time (Supplementary Fig. 1). 
%The WTA spiking neural network can converge to a good inference result in one second.

\begin{figure}[t]
	\begin{center}
		\includegraphics[width=\textwidth]{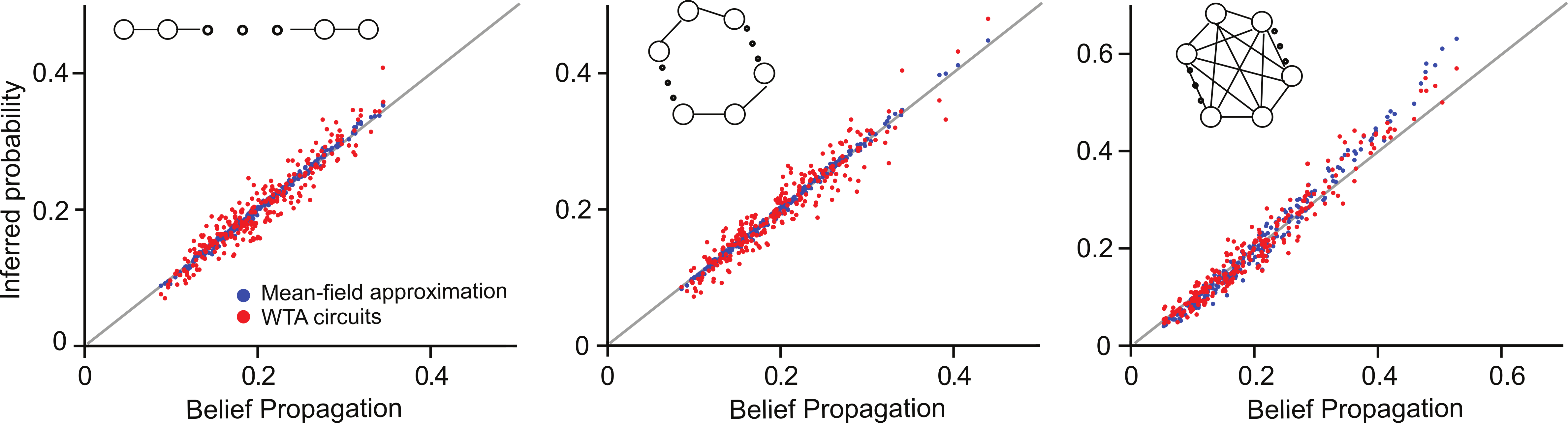}
	\end{center}
	\caption{Inference performance matched with mean-field inference algorithm and belief propagation in different graph topologies. All graphs have 10 nodes with topology as chain (left), single loop (middle), and fully connected graph (right). The performance of WTA circuits (red) and mean-field approximation (blue) is comparable to belief propagation. 
	} \label{fig:result2}
\end{figure}

Then we investigate whether the inference framework can be scaled up to more complex MRFs. Here we infer the marginal probability of multiple MRFs with different graph topologies as chain, single loop, and fully connected graph. As in Figure \ref{fig:result2} for the graphs with 10 nodes, one can find that the WTA circuits can get comparable results as belief propagation algorithm and mean-field approximation. 

Note that for a fullyc connected graph, there is a tendency that mean-field approximation are moving to zero and one for marginal probability. Such a phenomena was observed previously, which shown the marginal probabilities obtained by mean-field approximation are overconfident than the one obtained by belief propagation \cite{weiss2001comparing}. However, our WTA inference is better in this sense. We further test this point with a full connected graph with 20 nodes (see Supplementary Fig. 2), indeed, our WTA inference is more close to the result of belief propagation than that of mean-field approximation.

\section{Discussion}
In this study, we prove that there exists an exact equivalence between the neural dynamics of a network consisted of WTA circuits and the mean-field inference of probabilistic graphical models. We show the WTA circuits are able to represent distribution and implement inference of arbitrary probabilistic graphical models.  Our study suggests that the WTA circuit can be seen as the basic neural network motif for probabilistic inference at the level of neuronal circuits. This may offer a functional explanation for the existence of a large scale of WTA-like neuronal connectivities in cortical microcircuits. 

Unlike many previous neural circuits proposed for probabilistic inference \cite{rao2004bayesian,Sch2006The,Steimer2009Belief,Litvak2009Cortical,George2009Towards,Friston2017Active}, where each population of neurons has different network topology to implement the different and complex computation, our model is consist of a set of simple basic neural network motifs, and each motif works in a simple style.  In such a way, our proposed neural implementation is plausible as most computations during the cognitive behaviors are very conserved in that different part of the brain and different modality of sensory processing seems to be shared for neural information computation \cite{shimaoka2018effects,saleem2013integration}.

Difference approaches can be used to approximate Bayesian inference, among of which there are belief propagation and mean-field approximation working in a typical fashion. It was shown that the marginal probabilities obtained by mean-field approximation are overconfident than that obtained by belief propagation \cite{weiss2001comparing}, which means the marginal probabilities of mean-field approximation are closer to zero and one than the truth marginal probabilities. This interesting observation suggests that our proposed model with WTA circuits is suitable for computation of probabilistic reasoning. In the end, the brain has to shift an attention, select an action, and make a decision in face of uncertainties \cite{Pouget2013Probabilistic,itti2001computational}. 

It remains unclear that how different setting-ups of network with more components included from neuroscience can affect the inference result under different methods. For instance, with a graph topology more similar to the neuronal network in some part of the human brain area, or some neuronal network from typical well studied animals \cite{Pouget2013Probabilistic, Bullmore2009Complex}, the computation of WTA circuits proposed here could be explored. Perhaps, a more powerful utility of WTA circuits could be demonstrated for probabilistic reasoning and inference of the brain.

\bibliographystyle{}
\bibliography{ref}

%References follow the acknowledgments. Use unnumbered first-level
%heading for the references. Any choice of citation style is acceptable
%as long as you are consistent. It is permissible to reduce the font
%size to \verb+small+ (9 point) when listing the references. {\bf
%  Remember that you can use more than eight pages as long as the
%  additional pages contain \emph{only} cited references.}
%\medskip
%
%\small
%
%[1] Alexander, J.A.\ \& Mozer, M.C.\ (1995) Template-based algorithms
%for connectionist rule extraction. In G.\ Tesauro, D.S.\ Touretzky and
%T.K.\ Leen (eds.), {\it Advances in Neural Information Processing
%  Systems 7}, pp.\ 609--616. Cambridge, MA: MIT Press.
%
%[2] Bower, J.M.\ \& Beeman, D.\ (1995) {\it The Book of GENESIS:
%  Exploring Realistic Neural Models with the GEneral NEural SImulation
%  System.}  New York: TELOS/Springer--Verlag.
%
%[3] Hasselmo, M.E., Schnell, E.\ \& Barkai, E.\ (1995) Dynamics of
%learning and recall at excitatory recurrent synapses and cholinergic
%modulation in rat hippocampal region CA3. {\it Journal of
%  Neuroscience} {\bf 15}(7):5249-5262.

\end{document}